\documentclass{ws-ijmpa}

\begin{document}

\markboth{Aleksandra Biegun for the $\overline{\textrm{P}}$ANDA collaboration}
{EMC studies using the simulation framework of $\overline{\textrm{P}}$ANDA}

%
\catchline{}{}{}{}{}
%

\title{EMC studies using the simulation framework of $\overline{\textrm{P}}$ANDA}

\author{Aleksandra Biegun for the $\overline{\textrm{P}}$ANDA collaboration 
\footnote{On leave of absence from the INP, Polish Academy of Science, Krak\'ow, Poland.} 
\footnote{E-mail address: a.k.biegun@rug.nl.}}

\address{Kernfysisch Versneller Instituut, University of Groningen, Zernikelaan 25, 9747 AA Groningen, The Netherlands}


\maketitle

\begin{history}
\received{Day Month Year}
\revised{Day Month Year}
\end{history}

\begin{abstract}
The Anti-{\bf{P}}roton {\bf{AN}}nihilation at {\bf{DA}}rmstadt 
($\overline{\textrm{\bf{P}}}$ANDA) experiment proposed at the 
{\bf{F}}acility for {\bf{A}}ntiproton and {\bf{I}}on {\bf{R}}esearch (FAIR) 
in Darmstadt (Germany) will perform a high-precision spectroscopy of charmonium
and exotic hadrons, such as hybrids, glueballs and hypernuclei. 
A highly intense beam of anti-protons provided
by {\bf{H}}igh {\bf{E}}nergy {\bf{S}}torage {\bf{R}}ing (HESR) with an unprecedented resolution
will scan a mass range of 2 to 5.5 GeV/c$^{2}$. 

In preparation for expe\-riments with $\overline{\textrm{P}}$ANDA, 
careful and large-scale si\-mu\-lation stu\-dies
need to be performed in the coming years to determine analysis strategies,
to provide feedback for the design, construction and performance optimisation of individual 
detector components and to design methods for the calibration and interpretation 
of the experimental results. Results of a simulation for the {\bf{E}}lectro{\bf{M}}agnetic {\bf{C}}alorimeter (EMC),
built from lead tungstate (PWO) crystals and placed inside the Target Spectrometer 
(TS), are presented.  
The simulations were carried out using the PandaRoot framework, which is based on ROOT 
and being developed by the $\overline{\textrm{P}}$ANDA collaboration.

\end{abstract}

\ccode{PACS numbers: 13.25.Ft, 87.64.Aa, 01.50.hv}

\section{Motivation}    

With the $\overline{\textrm{P}}$ANDA experiment at FAIR in Darmstadt\cite{PANDAtpr2005}, 
a high resolution hadron spectroscopy will be performed. 
The charmonium states partly discovered with e\-xis\-ting $e^{+}e^{-}$ experiments
will be measured with the $\overline{\textrm{P}}$ANDA with much higher re\-so\-lution, 
which could not be achieved by $e^{+}e^{-}$ machines, but only via $\bar{p}p$ processes.
The exotic hybrid and glueball states, predicted by lattice QCD, as well as $\bar{p}A$ 
collisions will be also investigated.
The experimental setup of the $\overline{\textrm{P}}$ANDA project needs 
to be able to reconstruct predicted states and their decay channels with high precision.
Both, charged and neutral decay products of resonances, have to be detected
with very good spatial and energy resolutions.
For this purpose, detailed simulations and a corresponding analysis of various physics channels are 
performed.  
In this paper, the decay of the the $\bf{h_{c}}$ charmonium state via the following reaction 
is discussed:
\begin{equation}
\overline{\textrm{p}}+{\textrm{p}} \rightarrow \bf{h_{c}} \rightarrow 
\eta_{c} + \gamma \rightarrow (\pi^0+\pi^0+\eta) + \gamma \rightarrow 7\gamma.
\label{reaction}
\end{equation}
This is an example of a neutral decay, which can be studied with
$\overline{\textrm{P}}$ANDA. 
The $\overline{\textrm{P}}$ANDA detection system will be the perfect instrument 
for this, since it is highly compact, versatile and has a 4$\pi$ coverage.

\section{$\overline{\textrm{P}}$ANDA detection system and PandaRoot framework}  

\begin{figure}[pbh]
\psfig{file=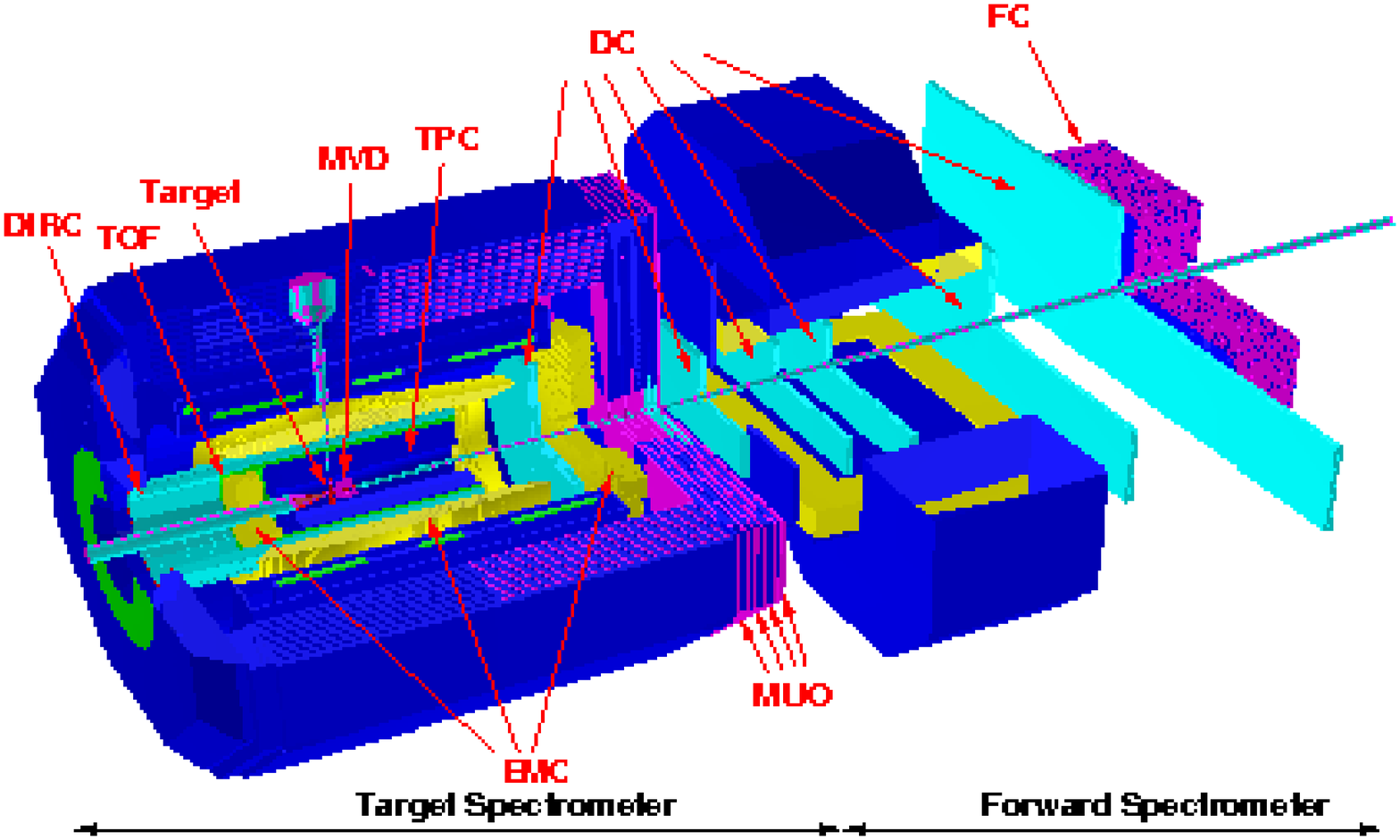,width=7.6cm}
\hspace{-0.4cm}\psfig{file=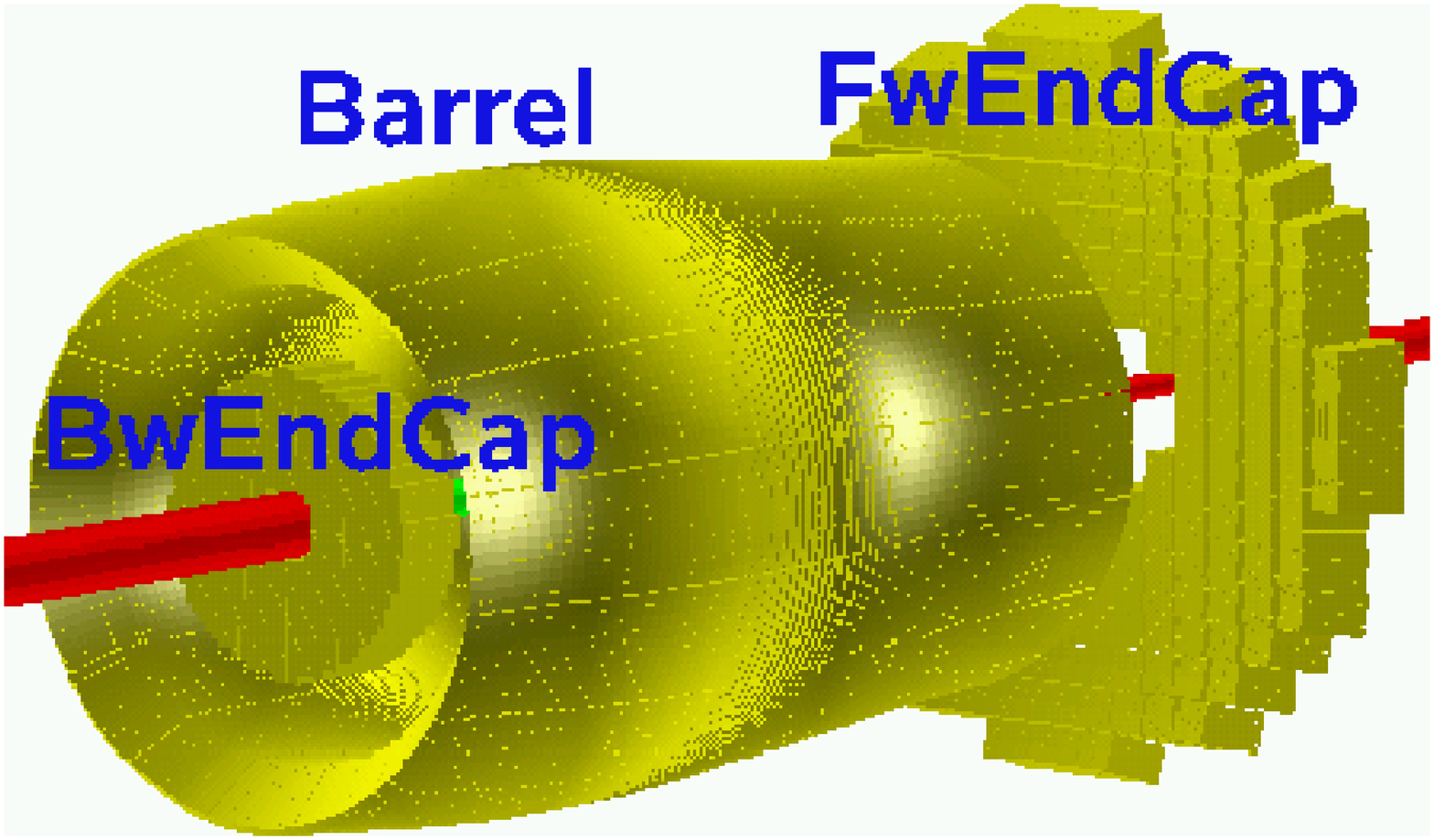,width=5.0cm}
\vspace*{8pt}
\caption{Left: The $\overline{\textrm{P}}$ANDA detection system. Right: The forward end-cap (FwEndCap), 
barrel (Barrel) and the backward end-cap (BwEndCap) of the EMC together with a beam pipe. Both
pictures are modelled using the PandaRoot simulation and analysis framework.\label{Panda_detector}}
\end{figure}

The $\overline{\textrm{P}}$ANDA detection system is presented in the left picture 
in Fig.~\ref{Panda_detector}.
It depicts the Micro Vertex Detector (MVD), Time Projection Chamber (TPC)
and, alternatively, Straw Tube Tracker (STT), ElectroMagnetic Calorimeter (EMC), Cherenkov detector (DIRC), 
 Muon detector (MUO), Time-Of-Flight (TOF), Drift Chambers (DC) and the Forward Calorimeter (FC). 
The EMC detector, placed inside the Target Spectrometer (TS), 
consists of the forward end-cap (FwEndCap), barrel (Barrel) and 
the backward end-cap (BwEndCap) and is shown in more detail on the right hand side in 
Fig.~\ref{Panda_detector}.
This highly granulated calorimeter, built from $\sim$16000 PWO crystals
with the size of about 2x2x20~$\textrm{cm}^{3}$ and 22~$\textrm{X}_{o}$ radiation length,
 was used to reconstruct the presented charmonium $\bf{h_{c}}$ (\ref{reaction}) from seven photons 
in the final state. 
\vspace*{-0.3cm}
\hspace{-0.5cm}Reaction (\ref{reaction}) has been simulated by using the \verb+EvtGen+\cite{EvtGen} 
event generator which has been included inside the simulation and analysis framework 
of the $\overline{\textrm{P}}$ANDA, called \verb+PandaRoot+\cite{SSpataro2008}.
\verb+EvtGen+, adapted for $\overline{\textrm{P}}$ANDA, 
was designed by BaBar\cite{BaBar} and originally used for the simulation of the physics of the B meson decays.
The \verb+Rho+ package\cite{KGoetzen2008} was applied to reconstruct invariant masses of 
final-state and inter\-me\-dia\-te particles of Eq.~(\ref{reaction}). 
The simulation framework is based on the Virtual Monte Carlo\cite{VMC2003} concept,
which allows to perform simulations for different transport
models without changing the user code or geometry description. 
The electronic response of the EMC crystals was simulated together with an optimised
cluster reconstruction analysis.
The EMC response has been compared and tuned to experimental data.

\section{Analysis of the $\bf{h_{c}}$ state}  

\begin{figure}[h]
\centerline{\psfig{file=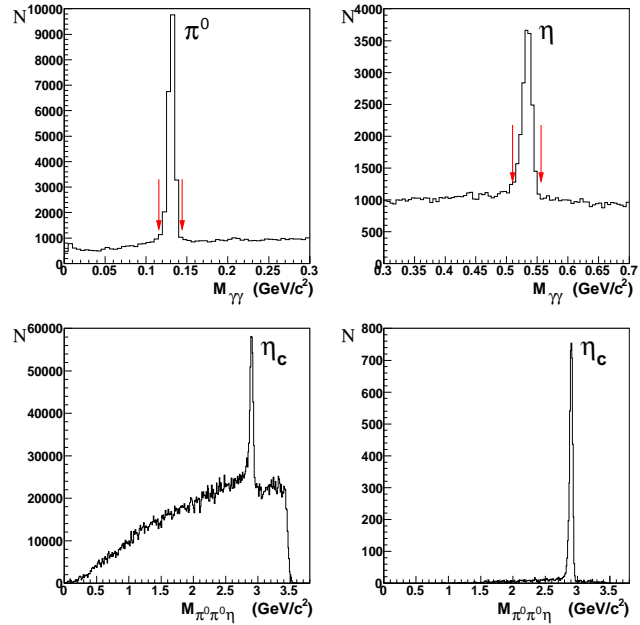,width=8.5cm}}
\vspace*{-2pt}
\caption{The invariant masses of $\pi^{0}$, $\eta$ and $\eta_{c}$ obtained from decay channel (\ref{reaction}). 
Lower left plot shows 
$\eta_{c}$ mass with a combinatorial background while the lower right one presents a very clean $\eta_{c}$ 
mass spectrum after cuts imposed at $\pi^{0}$ and $\eta$ masses showed in upper plots in this 
figure.\label{inv_masses}}
\end{figure}

The two-photon invariant mass spectrum, as presented in the top panels in Fig.~\ref{inv_masses}.
shows a clear signals from $\pi^{0}$ and $\eta$ particles on top of a continuous combinatorial
background.
\vspace*{-0.5cm}
The spectrum was fitted using a Gaussian function representing signal together with 
the second order polynomial function, representing the background.
A standard deviation, $\sigma$, of 6~MeV and 15~MeV was found for the $\pi^{0}$ and $\eta$,
respectively.
Pions and etas were selected by applying a windows of 3$\sigma$ around the two peaks 
as indicated by the arrows in both panels.
The 3$\sigma$ cuts have been applied to reduce the huge combinatorial background.
The bottom-left panel in Fig.~\ref{inv_masses}. depicts the invariant mass of the ($\pi^{0}$$\pi^{0}$$\eta$)
system for which all two-photon candidates were used to identify  $\pi^{0}$ and $\eta$ particles.
The bottom-right panel depicts the same analysis with the cuts applied to identify $\pi^{0}$ and 
$\eta$ mesons.
\begin{figure}[ht]
\begin{center}
\begin{minipage}[t]{0.48\textwidth}
\centerline{\epsfysize 4.6 cm
\epsfbox{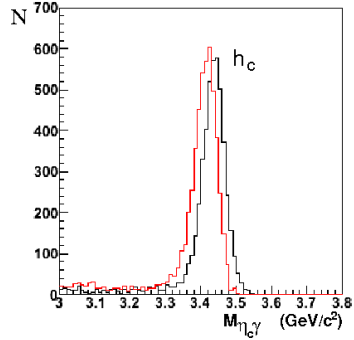}}
\end{minipage} \hfill
\begin{minipage}[t]{0.48\textwidth}
\centerline{\epsfysize 6.4 cm}
\vspace*{-100pt}
\caption{\label{hcinv_masses}The invariant mass of the charmonium $\bf{h_{c}}$ state, 
reconstructed from seven photons in the final state from the decay channel (\ref{reaction}). 
The solid line represents the spectrum obtained using GEANT3 while the dashed line represents a simulation using GEANT4.}
\end{minipage}
\end{center}
\end{figure}
Note that the peak to background ratio drastically improved by a factor 2500, as expected.
The combination of the reconstructed $\eta_{c}$ mass, 
together with the remaining photon, provides the identification of the $\bf{h_{c}}$ particle, 
as shown in Fig.~\ref{hcinv_masses}. 
In this analysis, only cuts for $\pi^{0}$ and $\eta$ masses have been applied.
The simulated spectrum representing the $\bf{h_{c}}$, yields to a peak-to-background of 30, 
a mass resolution (FWHM) of 70~MeV, and 
an efficiency of the $\bf{h_{c}}$ reconstruction of about 30\%. 
The distributions predicted by the two transport models, GEANT3 and GEANT4, are comparable.
A small shift of 25~MeV can be observed,
which can originate from small differences in the energy response between 
GEANT3 and GEANT4. A further study is in progress.

\section{Summary}

The analysis of the charmonium $\bf{h_{c}}$ state via the neutral channel 
has been done for two different transport models within the \verb+PandaRoot+ framework.
A more extensive simulation for other channels, including charged particles,
will be studied in the near future.

\section*{Acknowledgements}
This research is supported by Veni-grant 680-47-120 from the Netherlands Organisation for Scientific Research (NWO),      
the University of Groningen and the Gesellschaft f\"ur Schwerionenforschung mbH (GSI), Darmstadt.

\end{document}